\documentclass[epj,final]{svjour}
\usepackage{graphics}
\begin{document}
\title{Nonmesonic decay of the $\Lambda$-hyperon\\
in hypernuclei produced by p+Au collisions}
%
%
%
\author{B. Kamys\inst{1} \and P. Kulessa\inst{2} \and H. Ohm\inst{2}
\and K. Pysz\inst{3} \and Z. Rudy\inst{1} \and H. Str\"oher\inst{2}
\and W. Cassing\inst{4}
}                     
%
%
%
%
\institute{ M. Smoluchowski Institute of Physics,
Jagellonian University, PL-30059 Cracow, Poland \and 
Institut f\"ur Kernphysik, Forschungszentrum J\"ulich,
D-52425 J\"ulich, Germany \and
H. Niewodnicza\'nski Institute of Nuclear Physics,
PL-31342 Cracow, Poland \and
Institut f\"ur Theoretische Physik, Justus Liebig Universit\"at Giessen,
D-35392 Giessen, Germany}
\date{Received: date / Revised version: date}
%
%
\abstract{
The lifetime of the $\Lambda$-hyperon for the nonmesonic decay
$\Lambda$N $\rightarrow$ NN 
has been determined by a measurement at COSY J\"ulich of the delayed fission 
of heavy
hypernuclei produced in proton - Au collisions at T$_p$=1.9 GeV.  
It is found that  heavy hypernuclei
with mass numbers A$\approx$(180$\pm$5) and atomic numbers
Z$\approx$(74$\pm$2) fission with a lifetime 
$\tau_{\Lambda}$=130 $\pm$ 13(stat.) $\pm$ 15(syst.) ps.
This value together with the results obtained for other heavy hypernuclei
in previous investigations indicates -- on the confidence level of 0.9 --
a violation of the phenomenological $\Delta$I=1/2 rule 
for the $\Lambda$N $\rightarrow$ NN transitions as known from
the weak mesonic decays of kaons and hyperons. 
} 
\maketitle
%
%
%

The free $\Lambda$-hyperon undergoes almost with 100\% probability
the mesonic decay, i.e.
$\Lambda \rightarrow \pi^- + p$ or $\Lambda \rightarrow \pi^o + n$.
The energy release ($\approx$ 40 MeV) and its sharing among pion and
nucleon implies (due to momentum conservation) 
that the energy of the nucleon is much 
smaller than the Fermi energy of nucleons in the nucleus.
Therefore, such a process is strongly suppressed in nuclei and a different  
type of the hyperon decay --  nonmesonic decay --  dominates  
for all but the lightest hypernuclei. This decay can be induced by neutrons
(n+$\Lambda \rightarrow$ n+n) or by protons (p+$\Lambda \rightarrow$ p+n)
with an energy release much higher than in the mesonic decay 
($\approx$ 180 MeV). The higher nucleon energy, due to an equal sharing 
of the energy among both nucleons, implies that  
this process is not blocked by the Pauli principle.

The nonmesonic $\Lambda$ - decay represents an example for the nonleptonic weak 
interaction of baryons
with a change of strangeness ($\Delta$S = 1)  and isospin 
($\Delta$I= 1/2 or 3/2). Thus it is an analogue of the weak nucleon - nucleon
interaction, but it involves a new degree of freedom, i.e.  strangeness.
The strong and Coulomb interactions preserve  strangeness
and therefore the weak interaction responsible for the nonmesonic decay 
is not masked by contributions from  
these two interactions.  Therefore the nonmesonic decay enables 
to study both  parity violating
and parity conserving amplitudes in contrast to the 
nucleon-nucleon
interaction, where the latter amplitudes are completely masked by strong
and Coulomb forces \cite{KIS87}.  

The standard model of electro-weak interactions favors neither  
$\Delta$I= 1/2 transitions nor the $\Delta$I= 3/2 transitions, but  
experimental investigations on the  properties of mesonic decays 
of kaons and hyperons lead to the obvious dominance of the $\Delta$I= 1/2
part (the so called $\Delta$I= 1/2 rule) \cite{DON86}.  
The question arises whether this is also the case
for the nonmesonic decay of the $\Lambda$-hyperon. 
Data from the nonmesonic decay of light hypernuclei,
which were used to test this hypothesis in the phenomenological
model proposed by Block and Dalitz \cite{BLO63}, are affected by too large errors
to solve this problem unambigously \cite{ALB00}.  Another possibility
for testing the validity of the $\Delta$I= 1/2 rule is an investigation
of the dependence of the hyperon lifetime  for the nonmesonic decay
on the mass of the hypernucleus in which the hyperon 
is embedded \cite{RUD99}.    
Such a test requires, besides the knowledge of the lifetimes of light  
hypernuclei, also the precise knowledge of the lifetimes for heavy  hypernuclei.
The existing experimental results on the lifetime of heavy hypernuclei, 
which have been produced  in
antiproton interactions with Bi and U nuclei \cite{ARM93}, agree within the 
errors with the data obtained in proton induced reactions
on these targets \cite{OHM97,KUL98}.  Experiments with electrons on Bi nuclei
\cite{NOG86,NOG87} lead to an order of magnitude longer lifetime, however,
one has to note that the detection conditions of these experiments
were not suitable for a measurement of such short lifetimes as quoted
in antiproton and proton experiments.

In the present note new results on the lifetime measurements are 
presented which were obtained at COSY J\"ulich in proton collisions with gold nuclei
at T$_{p}$ = 1.9 GeV.
The details of the experimental apparatus and the data analysis are 
described elsewhere \cite{PYS99}. We briefly recall the physics of the
measurement and the detection principle. 
%
%
%
%
%
%
%
\begin{figure}
\resizebox{0.5\textwidth}{!}{%
\includegraphics{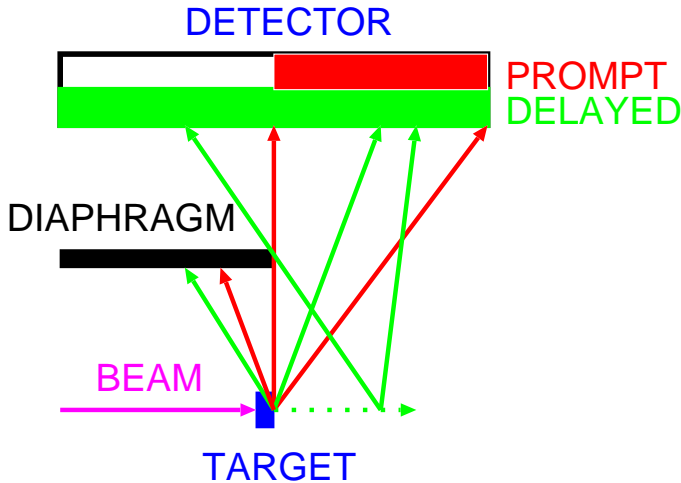}
}
\caption{The schematic view of the apparatus and the idea of the measurement.}
\label{fig:1}       
\end{figure}
The interaction of the proton beam with an energy above the (p,K$^+$) 
threshold with heavy target nuclei    
causes prompt fission of target nuclei as well as the associated (K$^+ \Lambda$)
production and for some fraction of reactions the production 
of hypernuclei.  The hypernuclei will promptly  
fission with large probability -- similarly to target nuclei -- or they can 
survive the prompt fission.  In the first case hypernuclei decay in the
target but the latter escape  and  
fission -- due to a rather long lifetime for the nonmesonic decay 
of the hyperon ($\approx 200 ps$) --
at some distance downstream of the target. 
Fragments from prompt fission of nuclei
and hypernuclei, which emerge from the target, can hit only that  
part of the detector which is not
shielded by a diaphragm, denoted
as "PROMPT" in  Fig. \ref{fig:1}.  
The fragments from the delayed fission of hypernuclei
are able to reach also the remaining part of the detector,  
not accessible by   
the prompt fission fragments.  Therefore the distribution of  hits
of the delayed fission fragments in the shadow region of the detector
is separated from the distribution of the prompt fission fragments
by a sharp edge and contains information on the lifetime 
of hyperons folded with the
velocity distribution of the hypernuclei 
(technique known as "recoil shadow method"). 

Such a method for the  separation of delayed fission fragments 
from the prompt fission \cite{MET74}
has been used in all experiments measuring the lifetime
of heavy hypernuclei \cite{ARM93,OHM97,KUL98,NOG86,NOG87,PYS99}. 
Experiments with a thin target in an  internal
proton beam have the advantage that  the recoiling hypernuclei
escape with larger velocities compared to hypernuclei
produced by electrons or antiprotons 
and therefore allow for the most accurate
determination of the lifetime.  However, it was observed 
in Ref. \cite{PYS99}
that the experiments in the internal beam are very sensitive to 
mechanical properties
of the targets, which must be very thin.  Especially the uranium target,  
which is the most efficient
for production of hypernuclei, is rather unstable both,  
in the form of UF$_4$ and UO$_2$.  It changes its shape 
during experiments causing rather large  background 
in the shadow region of the detector. 
By contrast in the present note we report on the experiment 
performed with a gold target, which is mechanically stable.
An additional advantage gives a favorable ratio
of the delayed fission events to the background appearing from 
prompt fission. This is illustrated in Table \ref{tab:1} , where
all factors influencing the ratio of the delayed fission cross section
to prompt fission cross sections are listed for U, Bi, and Au.  
 
%
%
\begin{table}
\caption{Comparison of calculated hypernuclei production cross sections
in proton induced reactions at T$_p$=1.9 GeV for 3 heavy nuclei 
($\sigma_{HY}$/$\mu$b),
the survival probability of the produced ('hot') hypernuclei 
against prompt fission ( P$_S$ ),
the probability of delayed fission of  ('cold') hypernuclei
induced by hyperon decay ( P$_{f_{\Lambda}}$ ),
the cross section for the delayed fission 
of hypernuclei ($\sigma_{del}$ / $\mu$b).  Also given are  
the cross sections for prompt fission 
of the target nucleus ($\sigma_{prompt}$ / b)
for U, Bi, and Au \cite{HUD76}-\cite{VAI81} .}

\label{tab:1}       
\begin{tabular}{llllll}
\hline\noalign{\smallskip}
   & \multicolumn{4}{c}{Theoretical values} & Exp. data \\ 
\hline\noalign{\smallskip}
Target & $\sigma_{HY}$/$\mu$b  & P$_S$  & P$_{f_{\Lambda}}$ & 
$\sigma_{del}$ / $\mu$b &
$\sigma_{prompt}$ / b \\
\noalign{\smallskip}\hline\noalign{\smallskip}
U & 410 & 0.12 & 0.85 & 42 & $\approx$1.5 \\ 
\noalign{\smallskip}\hline
Bi & 350 & 0.90 & 0.09 & 25 & $\approx$0.25 \\
\noalign{\smallskip}\hline
Au & 330 & 0.99 & 0.05 & 16 & $\approx$0.10 \\
\end{tabular}
\end{table}
The theoretical quantities presented in Table \ref{tab:1}  
have been evaluated in the coupled
channel Boltzmann-Uehling-Uhlenbeck approach for the first, 
fast stage of the reaction, 
accompanied by the statistical model
 for the second, slow stage of the reaction.
The last column contains experimental data taken from the literature  
\cite{HUD76,BOC78,VAI81}.
As can be seen, the decay rate of the 
delayed fission for Au targets is expected to be 
$\approx$3 times smaller
than that for U targets. 
However, the lower statistics for Au targets can be compensated to some
extent by a smaller background from the prompt fission fragments, 
because the cross section for prompt fission of Au nuclei by  protons
at 1.9 GeV energy is $\approx$16 times smaller than 
for a U target.

In the present experiment a 30 $\mu$g/cm$^2$ thick Au target on  
26 $\mu$g/cm$^2$ carbon backing was  irradiated by the 
internal proton beam of COSY 
with 5 $\cdot$ 10$^{10}$ protons circulating in the ring. 
The measurements were done
at 1.9 GeV -- to observe the decay of hypernuclei, and at 1.0 GeV 
-- to determine the background 
(the latter energy is low enough so that the production of hypernuclei 
is negligible).
COSY has been operated
in the so called supercycle mode \cite{PYS99},  in which the machine was 
switched between the
two energies every $\approx$ 18 s. Thus the  
properties of the target were the same for both energies. 
Other details of the experimental apparatus and the data analysis 
were the same
as  described in Ref. \cite{PYS99}.
%
%
\begin{figure}
\resizebox{0.45\textwidth}{!}{%
\includegraphics{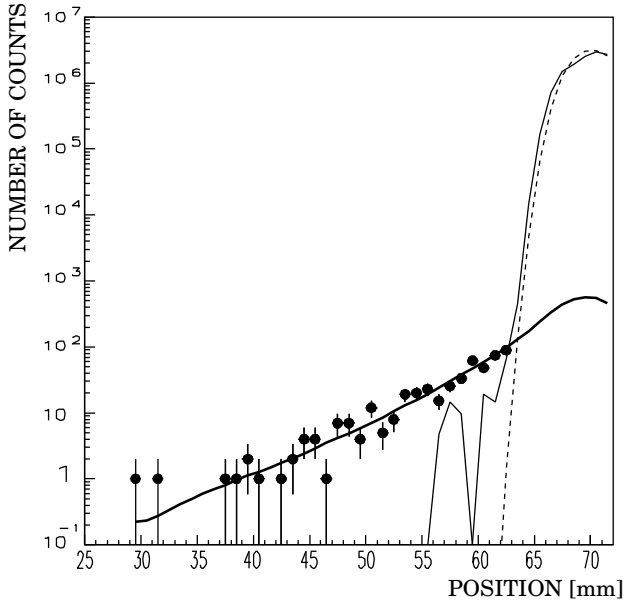}
}
\caption{The position distribution of hits of fission fragments
in position sensitive detectors. Details of the figure are discussed 
in the text below.}
\label{fig:2}       
\end{figure}

The projections of the two dimensional position distributions 
of the hits of  delayed
and prompt fission fragments in the multiwire proportional chambers
along the beam direction are presented 
in  Fig. \ref{fig:2}. The dots with error bars represent
the distribution measured at the proton energy T$_p$=1.9 GeV after
subtraction of the distribution measured at T$_p$=1.0 GeV (normalized
to the maximum of distributions). 
The latter distribution is
shown in Fig. \ref{fig:2} by the thin solid line.  
The data presented as full dots were used to extract the lifetime  
of the $\Lambda$ hyperon by a fit of the simulated  distribution
of delayed fission fragments - the thick solid line.
The dashed line in Fig. \ref{fig:2} corresponds to the simulated 
prompt fission fragment distribution 
(which is the same for T$_p$=1.9 GeV, 
and for 1.0 GeV). 
The velocity distribution of hypernuclei -- necessary for the simulation
of the delayed fission fragment distributions -- 
has been evaluated  
within the theoretical formalism mentioned above.  The reliability of the
calculations has been checked by a comparison of the experimental
and theoretical momentum distributions of kaons produced together
with the $\Lambda$-hyperons (in the associated production) and by
a comparison of experimental and theoretical momentum distributions
of the fragments from the proton induced prompt fission of the U target 
\cite{RUD98}.

The lifetime  extracted from the fit 
to the experimental data is:
\begin{eqnarray*}
\tau_{\Lambda} & =  & 130 \pm 13 (stat.) \pm 15 (syst.)\,\, ps. 
\end{eqnarray*}
%
%
%
This result agrees with the outcome of the experiment p+Bi   
(161 $\pm$ 7 (stat.) $\pm$ 14 ps) \cite{KUL98},
but is not in agreement with  
the published data for p+U reaction from Ref. \cite{OHM97} 
(240 $\pm$ 60 ps).  We point out, however, 
that a later reanalysis of the uranium data \cite{ZYC99},
in which Poisson statistics  
of events in position distributions was used instead  
the Gaussian statistics, lead to a  smaller value of the lifetime
(194 $\pm$ 55 ps).  Thus, the present value for p+Au agrees 
within the limits of  errors with the reanalyzed p+U data.
This also holds true for antiproton
induced hypernuclei production on Bi (180 $\pm$ 40 (stat.) $\pm$ 60 ps)
and U targets (130 $\pm$ 30 (stat.) $\pm$ 30 ps) \cite{ARM93}.
All these published data are biased with large errors -- with
the exception of the p+Bi experiment. The present p+Au 
experiment provides a new value for the lifetime of heavy hypernuclei, 
measured with a similar accuracy as that in the  p + Bi experiment.

 It is known \cite{RUD99}, that the shape of the mass dependence 
of the lifetime of hypernuclei is sensitive 
to the ratio R$_n$/R$_p$ of the neutron induced to proton induced $\Lambda$ 
nonmesonic decays, 
which results from the isospin structure of
the decay amplitudes.  
The absolute scale of the lifetimes is fixed by their values
for light hypernuclei, e.g. $^{11}$B, $^{12}$C, where lifetime is independent
of the R$_n$/R$_p$ ratio.  Thus, a validity of the phenomenological $\Delta$I = 1/2 rule, 
which implies that R$_n$/R$_p \le 2$,  put constraints on the lifetimes of heavy
hypernuclei.  According to this rule
the lifetime of heavy hypernuclei 
should  be larger than $\approx$ 180 ps for mass numbers A $\approx$ 180 .
The present experiment shows  
that the lifetime of these heavy hypernuclei is significantly shorter.
Thus, it indicates (together with p+Bi data) 
that the phenomenological rule
claiming that strange particles decay only  with
the change of isospin equal 1/2 is violated in the nonmesonic
decay of $\Lambda$ hyperons.  The confidence level  for this statement has been
determined - following the procedure given in Ref.\cite{RUD99} - to be equal 
$\approx$ 0.9.

\vspace{0.8cm}
\textbf{Acknowledgements}\\

The success of the experiment relied very much on the high quality of the 
Au targets which were prepared by Dr B. Lommel and her target laboratory
at GSI Darmstadt, Germany.

We are indebted to Prof.  O.W.B. Schult for stimulating discussions and
interest in these investigations.
 
The project has been supported by the DLR International Bureau of the BMBF,
Bonn, and the Polish Committee for Scientific Research (Grant No. 2P03B 16117).

%

\end{document}